# Giant Magnetothermal Conductivity Switching in Semimetallic WSi$_2$ Single Crystals

*Karl G. Koster,[1] Jackson Hise,[2] Joseph P. Heremans,[2,3] Joshua E. Goldberger[1]\**

[1]Department of Chemistry and Biochemistry, The Ohio State University, Columbus, OH 43210 USA

[2]Department of Mechanical and Aerospace Engineering, The Ohio State University, Columbus, OH 43210 USA

[3]Department of Physics, The Ohio State University, Columbus, OH 43210 USA

ABSTRACT:

Materials able to rapidly switch between thermally conductive states by external stimuli such as electric or magnetic fields can be used as all-solid-state thermal switches and open a myriad of applications in heat management, power generation and cooling.  Here, we show that the large magnetoresistance that occurs in the highly conducting semimetal α-WSi$_2$ single crystals leads to dramatically large changes in thermal conductivity at temperatures <100 K. At temperatures <20 K, where electron-phonon scattering is minimized, the thermal conductivity switching ratio between zero field and a 9T applied field can be >7. We extract the electronic and lattice components of the from the thermal conductivity measurements and show that the Lorenz number for this material approximates the theoretical value of L$_0$. From the heat capacity and thermal diffusivity, the speed of thermal conductivity switching is estimated to range from 1 x 10$^{-4}$ seconds at 5 K to 0.2 seconds at 100 K for a 5-mm long sample. This work shows that WSi$_2$, a highly conducting multi-carrier semimetal, is a promising thermal switch component for low-temperature applications such cyclical adiabatic demagnetization cooling, a technique that would enable replacing [3]He-based refrigerators.

INTRODUCTION

Compared to the extensive research focused on transistor materials in which the electronic conductance can be actively switched between various states, there has been considerably less effort focused on designing materials, systems, and concepts whose thermal conductance can be actively switched to enable the control of heat flow through a device.[1] All-solid-state heat switches operating fast and over large temperature ranges, if they existed, would enable new, durable, wear and maintenance-free approaches to thermal management.[2-3] In particular, thermal switches combined with heat accumulators can increase the thermal efficiency of thermodynamic power generation cycles in which the source of heat is transient.[1]

Thermal conductivity switches enable new schemes to temperature-control of components, such as batteries[4] and spacecraft,[5] under varying thermal loads. Heat switches are necessary components for electrocaloric and magnetocaloric solid-state cycles, in which no fluid is circulated but heat must be supplied to or drained from the active material.[6-8] In these applications, the switches have to function over the temperature range of the cycle and during transient operation such as cooldowns, a requirement that can be difficult to meet with switches based on phase changes in a material. Since the rest of the cycle is all solid-state with no moving parts, having a switch that shares those properties makes for durable maintenance-free operation of the whole heat engine.

Here, we focus on materials whose thermal conductivity changes under applied magnetic fields at temperatures below 10K. This temperature and field range matches that needed for continuously operating cyclical adiabatic demagnetization refrigerators (ADR's) that operate below 2K, using the same magnetic field for the magnetocaloric material (a paramagnet like gadolinium gallium garnet $Gd_3Ga_5O_{12}$) as for the switch. ADR's enable electronic loads to reach sub-Kelvin temperatures without the use of $^3He$, a rare and strategic gas. Continuously-operating ADR's are preferred over one-shot ones that can operate with He exchange gas as a switch, as occurs in dilution refrigerators. Heat switches are characterized by their switching ratio, the ratio between the conductance or conductivity of the switch in its "on" and the "off" state switching ratios (SR) $SR \equiv \kappa_{on}/\kappa_{off}$. Model calculations[9] of cyclical ADR's using metallic heat switches that have a $SR$ of the order of 3 to 10 indicate that a reasonable cooling power of (6 to 12 µW/gram of Gadolinium gallium garnet) can be achieved between 2K and 400 mK, albeit with a low coefficient of performance.

One possible mechanism to actively change the thermal conductivity of a material is by exploiting the electrical magnetoresistance of a material whose primary carriers of thermal conductivity are electrons. In a nonmagnetic metal the total thermal conductivity ($\kappa_{total}$) comes from the electronic component ($\kappa_e$), while the remainder comes from the phonon contribution to thermal conductivity, in which lattice vibrations conduct heat through the material ($\kappa_l$):

$$\kappa_{total} = \kappa_e + \kappa_l \qquad (1)$$

Each electron that travels through the material carries a charge $e$ and an amount of heat $k_BT$, so that if the scattering mechanisms that limit the mean free path for momentum and energy transport are the same, the Wiedemann-Franz[10] relationship holds between electrical conductivity and thermal conductivity:

$$\kappa_e = L \cdot T \cdot \sigma \qquad (2)$$

Here $T$ is the temperature, $\sigma$ is the electrical conductivity, and $L$ is the Lorenz ratio, which for free electrons is $L_0 = 2.44 \ 10^{-8}$ V²m⁻². To maximize the switching ratio between the "on" and the "off" state of the switch ($SR \equiv \kappa_{on}/\kappa_{off} = \kappa_{total}(0)/\kappa_{total}(B)$ ) of a magnetoresistive switch, one must have a very metallic material, where $\kappa_e > \kappa_l$; this is because the lowest value achievable for $\kappa_{total}(B)$ is $\kappa_l$. Because the range of $\kappa_l$ in most materials lies between ~0.1 to 1000 W m$^{-1}$ K$^{-1}$ from 1 to 300 K, the Wiedemann Franz law predicts that the electronic resistivity would have to range from $10^{-5}$ to $10^{-9}$ $\Omega$ m to meet the condition $\kappa_e = \kappa_l$ in the high conductance state. At the same time the material must have a large magnetoresistance, $MR = [\rho(B) - \rho(0)]/\rho(0)$. These two requirements are mutually counter-indicated in most materials, since a high MR comes from a high mobility, and good metals in which $\kappa_e > \kappa_l$ have much lower mobilities than semiconductors because their large Fermi surfaces result in intensive electron-phonon scattering.

Three classes of solutions have been proposed to circumvent this dilemma. The first is to use the giant magnetoresistance in magnetic multilayers such as Co/Cr, in which a cross-plane SR value of ~1.8 gas been observed.[11] The second is to use superconducting thermal switches. Superconductors, such as Nb[12], subject to a magnetic field greater than their critical field, become normal metals leading to an appreciable reduction in thermal $\kappa_e$. Superconducting Sn wires, In switches and Zn-foil switches have been made and studied[13] but except at temperatures below 100 millikelvin, their $SR$'s are relatively modest; in all cases, switching can only occur at temperatures below the superconducting transition temperature. The third solution is to use topological Weyl and Dirac semimetals including PtSn$_4$,[14] Bi$_x$Sb$_{1-x}$[15] and NbP.[16] The condition $\kappa_e > \kappa_l$ limits the $SR$ in all these solutions.

Non-magnetic, non-superconducting, non-topological semimetallic materials, like Bi,[17-18] Cd,[19-20] Ga,[19] or W[21] that have similar carrier densities and mobilities of both holes and electrons also undergo an appreciable magnetoelectrical resistance.[22] This large, non-saturating MR in semimetals comes from electron-hole compensation that gives rise to a mismatch between the Lorentz force and the Hall voltage.[6] Furthermore, when both electrons and holes are present in clean compensated semimetals, the ambipolar thermal conductivity can lead to an increase in $L_0$ over the free electron value, which is beneficial for the *SR*. Thus, fully evaluating the magnetothermal conductivity properties in semimetallic materials beyond the simple metals is essential to understand their potential and limitations.

Here we explore the magnetothermal conductivity switching in the compensated semimetal $WSi_2$, a material recently established to exhibit axis-dependent conduction polarity. Recent studies have found that at 2 K, certain crystals of $WSi_2$ exhibit a high transverse electronic MR of 4,000-8,000% at 14 T, where %MR = 100%*$[\rho(B)-\rho(0)]/\rho(0)$.[23-24] This large increase in electrical resistivity in the presence of a magnetic field seen in $WSi_2$ is expected to lead to a large increase in magnetothermal resistance. To investigate the extent to which the thermal conductivity of $WSi_2$ can be manipulated using an applied magnetic field, we evaluated the changes in electrical and thermal conductivity in two different single crystals from 2-300K and 0-9T. The total thermal conductivity can be significantly reduced upon application of a 9T magnetic field below 20 K, leading to a thermal conductivity switching ratio (SR) *SR* > 7. The thermal conductivity switching ratio is larger in the crystal with fewer defects, on account of its lower baseline electronic resistivity and larger corresponding magnetoresistance.

Negligible differences in the thermal switching ratio or temperature dependence are observed along the [100] or [001] directions. Both *MR* and *SR* decrease with increasing temperature, although both crystals retain an *SR* close to 1.1 at room temperature. We show that the Lorenz number for this material closely matches the free electron value, thereby indicating minimal ambipolar effects that contribute to thermal conductivity. We also calculate the thermal response time from the specific heat and thermal diffusivity, and show that for a 5 mm$^3$ sample, switching can occur in less than a second at room temperature and much faster at lower temperatures. Overall, this work shows that ultraclean, compensated semimetallic compounds such as $WSi_2$ show great promise as heat switches.

## EXPERIMENTAL PROCEDURES

WSi2 crystals were grown using a Xenon Optical Floating Zone Furnace using a previously described procedure.[23] Crystals were cut into 1 mm x 1 mm x 5 mm bars along the [100] and [001] directions using a diamond wire saw to maximize sample resistance, and these samples were polished using 3 mm diamond grit in oil and 0.05 mm alumina slurry on a precision lapping and polishing machine to minimize contact resistance. Contacts to the surface of the crystal were made using Transene GE-40 gold-epoxy paste. Transverse magnetoresistance measurements were performed using a 4-probe geometry in a Quantum Design 14T PPMS from 2-300K and from 0-9T using the DC resistivity option. Thermal conductivity measurements were performed from 2-300K and from 0-9T using the thermal transport option (TTO) in a Quantum Design 9T PPMS.

## RESULTS AND DISCUSSION

α-WSi₂ crystallizes into a body-centered tetragonal I4/*mmm* crystal structure (Figure 1a). This material has a metallic band structure with unique band curvatures that give rise to simultaneous electron and hole conduction along different crystallographic axes (Figure 1b and c).[4,5] Single crystals of α-WSi₂ previously synthesized using an optical floating zone furnace[4] were selected to study how the thermal conductivity of WSi₂ changes in the presence of an applied magnetic field. These single crystals were confirmed to be phase pure with single-crystal and powder X-ray diffraction, and were oriented along their primary [001] and [100] crystallographic directions using Laue backscattering X-ray diffraction.[4] These rod-shaped crystals were cut into 1 mm x 1 mm x 5 mm bars to maximize sample resistance for electrical measurements, with the crystallographic direction of interest ([100] or [001]) oriented along the length of the bar (Figure 1d). The surfaces of the bars were polished to minimize electrical contact resistances in the measurements.

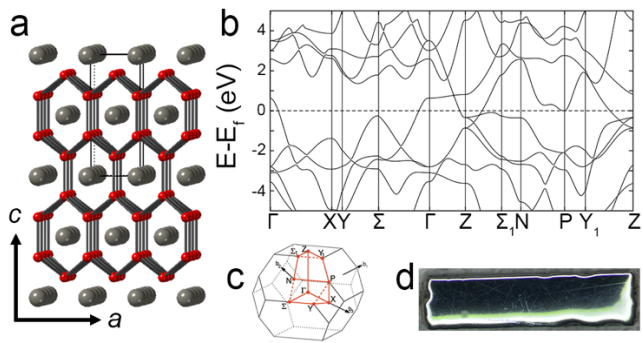

**Figure 1.** (a) The crystal structure of I4/*mmm* α-WSi₂. W atoms are shown in gray and Si atoms are shown in red. The unit cell is drawn in black. (b) The electronic band structure of WSi₂. (c) The Brillouin zone for a body-centered tetragonal material with *c>a*. (d) A photograph of a polished WSi$_2$ sample (5 mm in length, 1 mm in width and thickness) that was used for magnetoresistance and magnetothermal conductivity measurements.

The thermal conductivity of these polished bars of WSi$_2$ were evaluated from 2 to 300K in the presence of transversely applied magnetic fields from 0 to 9T using a steady-state method (Figure 2). At low temperatures (<100K), the thermal conductivity significantly decreases in the presence of a magnetic field. For crystal 1, the thermal conductivity at 2K decreased from 37.9 to 11.1 W m$^{-1}$ K$^{-1}$ in the [100] direction (Figure 2a) while it decreased from 43.5 to 13.7 W m$^{-1}$ K$^{-1}$ in the [001] direction (Figure 2b) when the magnetic field was ~~is~~ increased from 0 to 9T. A switching factor of $\kappa$(0T)/$\kappa$(9T) was determined from the thermal conductivity values measured at 0 and 9T at each temperature point, (Figure 2c). This factor is highest at 2K, where $SR$=$\kappa$(0T)/$\kappa$(9T) = 3.4 along the [100] direction while $SR$ = 3.2 along the [001] direction. For crystal 2 along [100], $SR$ = 7±0.1 at 7K, and its increase with decreasing temperature is not quite saturated yet. These are remarkable values for a metal.[8] $SR$ approaches unity at 300K.

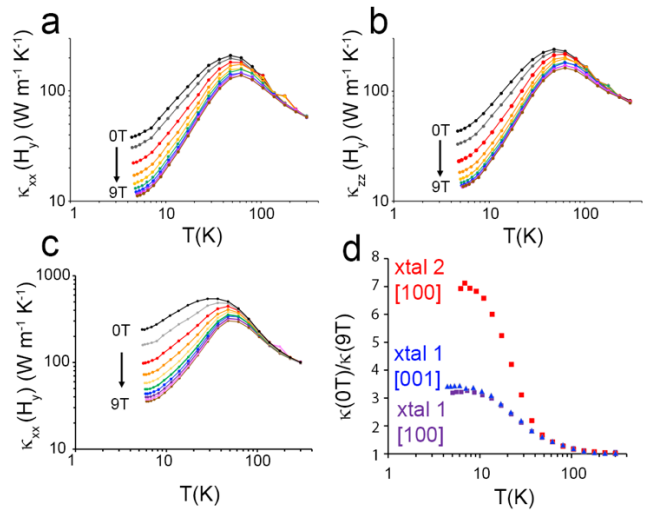

**Figure 2.** Thermal conductivity versus temperature with an applied transverse magnetic field of 0 to 9 T in 1 T increments for the (a) [100] and (b) [001] directions in crystal 1, and the (c) [100] direction in crystal 2. (d) The thermal conductivity switching factor, calculated as SR=k(0T)/k(9T).

To confirm that this reduction in thermal conductivity comes from the large magnetoelectrical resistance observed in WSi₂, we measured the electrical resistivity of the two WSi₂ crystals from 2-300K when transverse magnetic fields of 0-9T were applied (Figure 3). Residual resistivity values (RRR) from the zero-field measurements ($RRR = \frac{\rho(300K,0T) - \rho(2K,0T)}{\rho(2K,0T)}$) for crystal 1 were 37.6 for the [100] direction (Figure 3a) and 26.9 for the [001] direction (Figure 3b), indicating good crystal quality. Crystal 2 had a much higher RRR of 101, indicative of fewer defects, measured along the [100] direction. The presence of fewer defects in crystal 2 was further evidenced by the lower baseline resistivities. The Magnetoresistance maximizes at low temperatures in both crystals. The %MR for crystal 1 reached 2200-2400% at 9T from 2-21 K for the [100] direction (Figure 3a) and 3200-3500% at 9T from 2-21K for the [001] direction (Figure 3b), closely matching previously published values.[23-24] Crystal 2 had a much larger %MR of ~9,000-10,000% at 9T from 2-21K for the [100] direction. Above 21K, the magnetoresistance effect starts to fall as electron-phonon scattering decreases the mobility of the carriers in the material at higher temperatures, as occurs in nearly all metallic materials.[25]

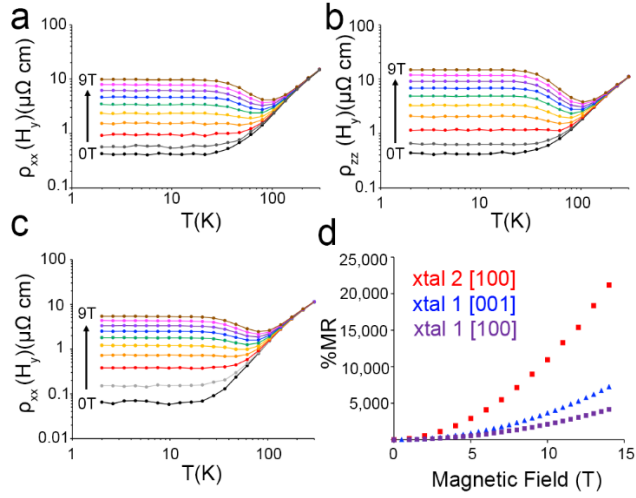

**Figure 3.** Electrical resistivity versus temperature with an applied transverse magnetic field varying from 0 to 9 T at 1 T increments for the (a) [100] and (b) [001] directions in crystal 1, and (c) [100] direction of crystal 2. (d) The %MR measured at 2K for crystal 1 along the [100] direction (purple), crystal 1 along the [001] direction (blue), and crystal 2 along the [100] direction (red).

The magnetic field-induced reduction in both thermal and electrical conductivities occur across similar temperature ranges (2-100K), with the effects being most pronounced at low temperature. Importantly, although the magnetoelectrical resistance in the material increases quadratically with magnetic field and does not saturate, the thermal conductivity decreases asymptotically as magnetic field is increased. This is because $\kappa_e$ is nearly eliminated, leaving only $\kappa_l$, which remains unchanged in the presence of a magnetic field (Figure 4a and b). The thermal conductivity data does not fully reach saturation at 9T, but is expected to do so at higher magnetic fields. To determine the value of $\kappa_l$ at each temperature point, the field-dependent thermal conductivity data for each temperature was fitted to the function

$\kappa_{total} = \kappa_l + \frac{\gamma}{1+\beta^2 B^2}$, in which $\gamma$ is a linear coefficient, $\beta$ is a quadratic coefficient representing the magnetoconductivity, and B is magnetic field. The last term in the equation corresponds to the change in thermal conductivity corresponding to the electrical magnetoresistance. The data could only be fit at temperatures up to 36K. Above this temperature, the much smaller magnetoresistance effect leads to a smaller change in thermal conductivity across all fields, introducing a large uncertainty in the determination of $\kappa_l$. Subtracting the fitted value of $\kappa_l$ from the experimentally determined $\kappa_{total}$, 0T yields a value for $\kappa_e$ at 0T, or $\kappa_{e,0T}$. $\kappa_{total}$, $\kappa_l$, and $\kappa_{e,0T}$ are all plotted together versus temperature up to 36K in Figure 4 c and d. $\kappa_e$ is also plotted from 2-36K and 0-9T in Figure 4 e and f, showing how the electronic component of the thermal conductivity decreases at higher field via the magnetoresistance mechanism. $\kappa_l$ remains unchanged while $\kappa_e$ decreases with increasing magnetic field, leading to an overall decrease in $\kappa_{total}$ when a magnetic field is applied.

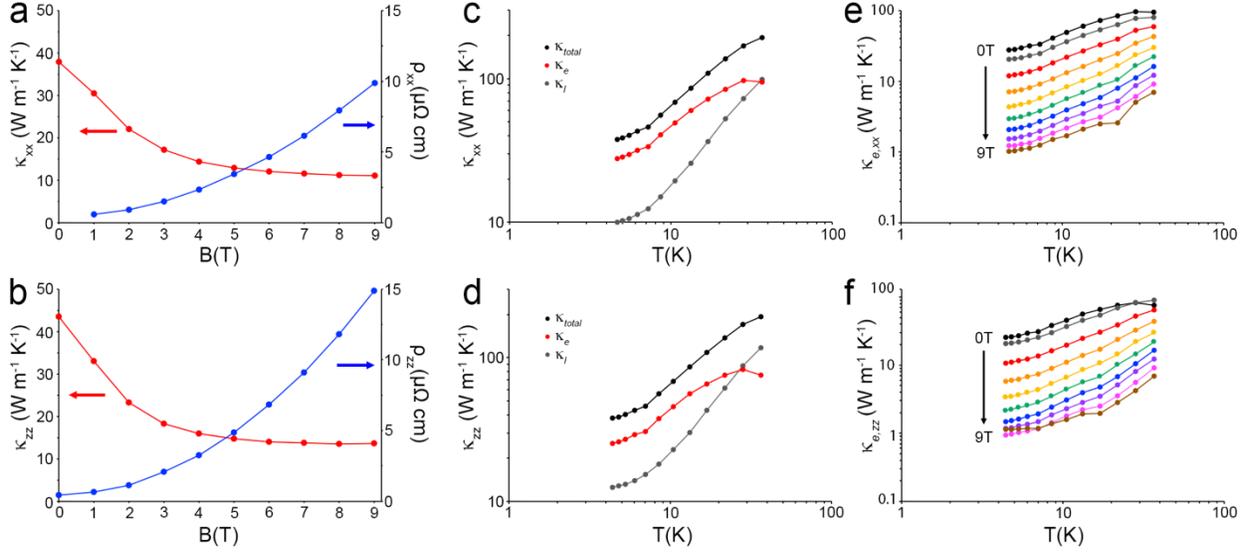

**Figure 4.** Thermal conductivity (red, left axes) and electrical resistivity (blue, right axes) versus magnetic field at 2K for the (a) [100] and (b) [001] directions for crystal 1. Plot of $\kappa_{\text{total}}$ (black), $\kappa_l$ (gray), and $\kappa_{e,0T}$ (red) versus temperature measured along the (c) [100] and (d) [001] directions for crystal 1. $\kappa_e$ plotted versus temperature for magnetic field values of 0-9T for the (e) [100] and (f) [001] directions for crystal 1.

Using the $\kappa_e$ values that we found by subtracting $\kappa_l$ from $\kappa_{total}$, we obtained a value for the Lorenz constant using the Wiedemann-Franz law, $L = \frac{\kappa_e}{T \cdot \sigma}$. Next, the ratio $L/L_0$ was calculated at fields from 0-9T and at temperatures from 2-36K to determine if magnetoresistance and magnetothermal conductivity change by the same proportion, which would indicate that this material follows the free-electron Wiedemann-Franz Law. The results for both the [100] and [001] directions are plotted in Figure 5. $L/L_0$ values for both crystallographic directions match closely between fields at each given temperature point and are close to 1 (0.6-1.2) indicating that the experimentally observed relationship between magnetothermal conductivity and magnetoelectrical conductivity is relatively proportional and obeys the Wiedemann-Franz Law. This proves that the large magnetoresistance in WSi$_2$ is what gives rise to its accompanying decrease in thermal conductivity.

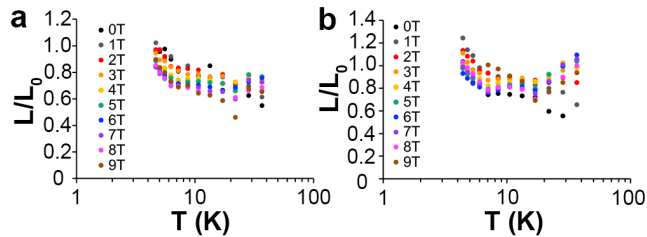

**Figure 5.** L/L$_0$ for the (a) [100] and (b) [001] direction under different applied magnetic fields for crystal 1.

A functional thermal switch should be capable of rapidly switching between the conductive and non-conductive states. This rate is limited by the thermal diffusivity of the material (given in m$^2$/s), with diffusivity equal to the thermal conductivity divided by the specific heat capacity: $\alpha = {\kappa}/{(\rho * c_p)}$, in which $\alpha$ is the diffusivity, $c_p$ is the specific heat capacity in J/(kg K) and $\rho$ is the material density in kg/m$^3$. The experimentally measured molar heat capacity ($c_{p,m}$) is plotted in Figure 6a and, along with the thermal conductivity data plotted in Figure 2, can be used to determine the thermal diffusivity of the material. Then, a thermal response time for a sample of a certain size is given by $\tau = \ell^2/\alpha$, in which $\tau$ is the time constant and $\ell$ is the sample length. This value of $\tau$ is given for a 5 mm-long sample of WSi₂ in Figure 6. $\tau$ values are nearly isotropic along both primary crystallographic directions, ranging from $1\times10^{-4}$ s at 2K to 0.8-1.1 s at 300K. Practically, this means that the thermal switching speed will ultimately be dictated by the speed in which a magnetic field can be applied to the material, which is much faster than our current measurement setup.

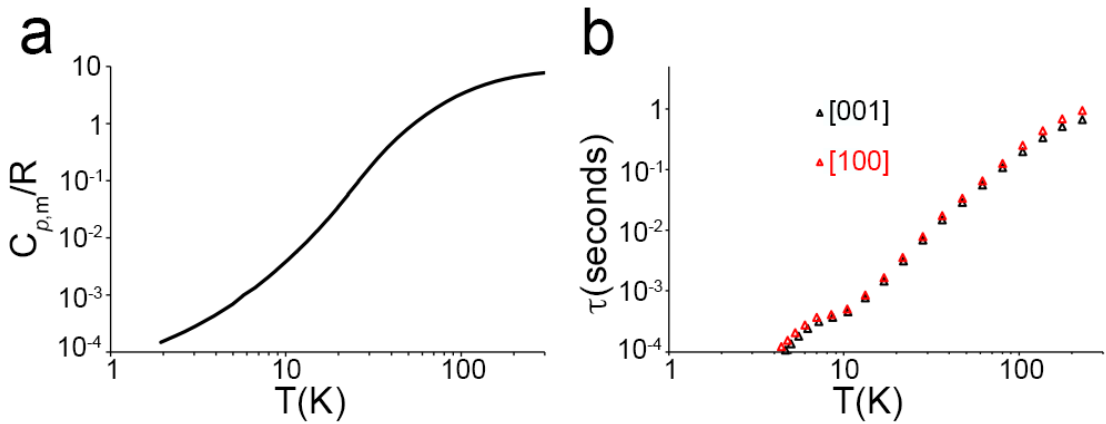

**Figure 6.** (a) Measured molar heat capacity of crystal 1. (b) The calculated thermal response time for a 5-mm long sample of WSi₂, shown for the [100] (red) and [001] (black) directions.

CONCLUSIONS

A large 9000-10000% magnetoresistance in WSi₂ at low temperatures of 2-21K and fields of 0-9T can be attributed to the electron-hole compensation in this two-carrier material and gives rise to an accompanying 7-fold decrease in the material's thermal conductivity, which closely follows the Wiedemann-Franz Law. Furthermore, a WSi₂ crystal with a larger relative resistivity ratio has a larger SR, indicating that minimization of defects during crystal growth will provide an opportunity to enhance the magnitude of thermal switching. The appreciable change in the thermal conductivity in the presence of a magnetic field, along with the material's ability to rapidly switch between thermally conductive and less conductive states, makes $WSi_2$ a promising material system for thermal switching applications at low temperatures.

AUTHOR INFORMATION

**Corresponding Author**

*goldberger.4@osu.edu

**Author Contributions**

The manuscript was written through contributions of all authors. All authors have given approval to the final version of the manuscript.

**ACKNOWLEDGMENT**

This work was primarily supported by the ONR MURI "Extraordinary electronic switching of thermal transport", grant number N00014-21-1-2377. JH is supported by NSF CBET grant "EAGER: CRYO: Development of a sub-kelvin refrigerator using magnetic field activated solid-state thermal switches based on thermal chiral anomaly", grant number CBET 2232811. The material used in this work was supported by the NSF-MIP Platform for the Accelerated Realization, Analysis, and Discovery of Interface Materials (PARADIM) under Cooperative

Agreement No. DMR-2039380. Partial funding for shared facilities used in this research was provided by the Center for Emergent Materials: an NSF MRSEC under award number DMR-2011876. The authors would like to acknowledge Lucas Pressley and Mojammel Khan at the NSF PARADIM facility for helping to grow large single crystals of $WSi_2$.

REFERENCES

1. Wehmeyer, G.; Yabuki, T.; Monachon, C.; Wu, J.; Dames, C., Thermal Diodes, Regulators, and Switches: Physical Mechanisms and Potential Applications. *Appl. Phys. Rev.* **2017,** *4*, 41304.

2. Liu, C. H.; Chen, Z. H.; Wu, C.; Qi, J.; Hao, M. L.; Lu, P.; Chen, Y. F., Large Thermal Conductivity Switching in Ferroelectrics by Electric Field-Triggered Crystal Symmetry Engineering. *ACS Appl. Mater. Interfaces* **2022,** *14*, 46716-46725.

3. Qian, X.; Zhou, J. W.; Chen, G., Phonon-Engineered Extreme Thermal Conductivity Materials. *Nat. Mater.* **2021,** *20*, 1188-1202.

4. Du, T.; Xiong, Z.; Delgado, L.; Liao, W.; Peoples, J.; Kantharaj, R.; Chowdhury, P. R.; Marconnet, A.; Ruan, X., Wide Range Continuously Tunable and Fast Thermal Switching Based on Compressible Graphene Composite Foams. *Nat. Commun.* **2021,** *12*, 4915.

5. Hengeveld, D. W.; Mathison, M. M.; Braun, J. E.; Groll, E. A.; Williams, A. D., Review of Modern Spacecraft Thermal Control Technologies. *HVAC&R Res.* **2010,** *16*, 189-220.

6. Klinar, K.; Vozel, K.; Swoboda, T.; Sojer, T.; Rojo, M. M.; Kitanovski, A., Ferrofluidic Thermal Switch in a Magnetocaloric Device. *iScience* **2022**, *25*, 103779.

7. Klinar, K.; Kitanovski, A., Thermal Control Elements for Caloric Energy Conversion. *Renewable Sustainable Energy Rev.* **2020,** *118*.

8. Silva, D. J.; Bordalo, B. D.; Pereira, A. M.; Ventura, J.; Araujo, J. P., Solid State Magnetic Refrigerator. *Appl. Energy* **2012,** *93*, 570-574.

9. Shirron, P., Optimization Strategies for Single-Stage, Multi-Stage and Continuous Adrs. *Cryogenics* **2014,** *62* 140-149.

10. Franz, R.; Wiedemann, G., Ueber Die Wärme-Leitungsfähigkeit Der Metalle. *Ann. Phys. (Berlin, Ger.)* **1853,** *165*, 497-531.

11. Kimling, J.; Wilson, R. B.; Rott, K.; Kimling, J.; Reiss, G.; Cahill, D. G., Spin-Dependent Thermal Transport Perpendicular to the Planes of Co/Cu Multilayers. *Phys. Rev. B: Condens. Matter Mater. Phys.* **2015,** *91*, 144405.

12. Wasim, S. M.; Zebouni, N. H., Thermal Conductivity of Superconducting Niobium. *Phys. Rev.* **1969,** *187*, 539.

13. Schuberth, E., Superconducting Heat Switch of Simple Design. *Rev. Sci. Eng.* **1984,** *55*, 1486-1488.

14. Fu, C. G.; Guin, S. N.; Scaffidi, T.; Sun, Y.; Saha, R.; Watzman, S. J.; Srivastava, A. K.; Li, G. W.; Schnelle, W.; Parkin, S. S. P.; Felser, C.; Gooth, J., Largely Suppressed Magneto-Thermal Conductivity and Enhanced Magneto-Thermoelectric Properties in PtSn$_4$. *Research* **2020**, *2020*, 4643507.

15. Vu, D.; Zhang, W. J.; Sahin, C.; Flatte, M. E.; Trivedi, N.; Heremans, J. P., Thermal Chiral Anomaly in the Magnetic-Field-Induced Ideal Weyl Phase of Bi$_{1-x}$Sb$_x$. *Nat. Mater.* **2021**, *20*, 1525-1531.

16. Gooth, J.; Niemann, A. C.; Meng, T.; Grushin, A. G.; Landsteiner, K.; Gotsmann, B.; Menges, F.; Schmidt, M.; Shekhar, C.; Suss, V.; Hune, R.; Rellinghaus, B.; Felser, C.; Yan, B. H.; Nielsch, K., Experimental Signatures of the Mixed Axial-Gravitational Anomaly in the Weyl Semimetal NbP. *Nature* **2017**, *547*, 324-327.

17. Alers, P. B.; Webber, R. T., The Magnetoresistance of Bismuth Crystals at Low Temperatures. *Phys. Rev.* **1953**, *91*, 1060-1065.

18. Uher, C.; Goldsmid, H. J., Separation of the Electronic and Lattice Thermal Conductivities in Bismuth Crystals. *Phys. Status Solidi B* **1974**, *65*, 765-772.

19. Mendelssohn, K. A. G.; Rosenberg, H. M., The Thermal Conductivity of Metals in High Magnetic Fields at Low Temperatures. *Proc. R. Soc. Lond.* **1953**, *A218*, 190-205.

20. Laudy, J.; Knol, A., A Cadmium Heat Switch. *Cryogenics* **1966**, *6*, 370.

21. Batdalov, A.; Red'ko, N., Lattice and Electronic Thermal Conductivities of Pure Tungsten at Low Temperatures. *Sov. Phys. Solid State* **1980,** *22*, 664–666.

22. Zhang, S.; Wu, Q.; Liu, Y.; Yazyev, O. V., Magnetoresistance from Fermi Surface Topology. *Phys. Rev. B* **2019,** *99*, 35142.

23. Koster, K. G.; Deng, Z.; Heremans, J. P.; Windl, W.; Goldberger, J. E., Axis-Dependent Conduction Polarity in $WSi_2$ Single Crystals. *Chem. Mater.* **2023,** *35*, 4228-4234.

24. Mondal, R.; Sasmal, S.; Kulkarni, R.; Maurya, A.; Nakamura, A.; Aoki, D.; Harima, H.; Thamizhavel, A., Extremely Large Magnetoresistance, Anisotropic Hall Effect, and Fermi Surface Topology in Single-Crystalline $WSi_2$. *Phys. Rev. B* **2020,** *102*, 115158.

25. Chung, D.-Y.; Hogan, T. P.; Rocci-Lane, M.; Brazis, P.; Ireland, J. R.; Kannewurf, C. R.; Bastea, M.; Uher, C.; Kanatzidis, M. G., A New Thermoelectric Material: $CsBi_4Te_6$. *J. Am. Chem. Soc.* **2004,** *126*, 6414-6428.